\documentclass[a4paper, longbibliography, twocolumn, superscriptaddress, showkeys, floatfix, nofootinbib]{revtex4-1}

\usepackage[utf8]{inputenc}
\usepackage[english]{babel}
\usepackage{amssymb}
\usepackage{amsmath}
\usepackage{graphicx}
\usepackage[caption=false]{subfig}
\usepackage{float}
\usepackage{nameref}
\usepackage[normalem]{ulem}
\usepackage{color}

\renewcommand{\subsubsection}[1]{}

\begin{document}

\author{Kaj-Kolja Kleineberg}
% \email{kkl@ffn.ub.edu}
\email{kajkoljakleineberg@gmail.com}
\author{Mari\'an Bogu\~{n}\'a}
\affiliation{Departament de F\'isica Fonamental, Universitat de Barcelona, 
Mart\'i i Franqu\`es 1, 08028 Barcelona, Spain}
\author{M. Ángeles Serrano}
\affiliation{Instituci\'o Catalana de Recerca i Estudis Avan\c{c}ats (ICREA), Passeig Llu\'is Companys 23, E-08010 Barcelona, Spain}
\affiliation{Departament de F\'isica Fonamental, Universitat de Barcelona, Mart\'i i Franqu\`es 1, 08028 Barcelona, Spain}
\author{Fragkiskos Papadopoulos}
\email{f.papadopoulos@cut.ac.cy}
\affiliation{Department of Electrical Engineering, Computer Engineering and Informatics, Cyprus University of Technology, 33 Saripolou Street, 3036 Limassol, Cyprus}
\title{Hidden geometric correlations in real multiplex networks}

\begin{abstract}
Real networks often form interacting parts of larger and more complex systems. Examples can be found in different domains, ranging from the Internet to structural and functional brain networks. Here, we show that these multiplex systems are not random combinations of single network layers. Instead, they are organized in specific ways dictated by hidden geometric correlations interweaving the layers. We find that these correlations are significant in different real multiplexes, and form a key framework for answering many important questions. Specifically, we show that these geometric correlations facilitate: (i) the definition and detection of multidimensional communities, which are sets of nodes that are simultaneously similar in multiple layers; (ii) accurate trans-layer link prediction, where connections in one layer can be predicted by observing the hidden geometric space of another layer;  and (iii) efficient targeted navigation in the multilayer system using only local knowledge, which outperforms navigation in the single layers only if the geometric correlations are sufficiently strong. Our findings uncover fundamental organizing principles behind real multiplexes and can have important applications in diverse domains.
\end{abstract}

\maketitle

%%%%%%%%%%%%%%%%%%%%%%%%%%%%%%%%%%%%%%%%%%%%%%%%%%%%%%%%%%%%%%%%%%%%%%%%%%%%%%%%%%%

Real networks are often not isolated entities but instead can be considered constituents of larger systems, called multiplexes or multilayer networks~\cite{multilayer:kivel, arenas:radicchi:2013, ginestra:natphys, radicci:percolation, arenas:multiplex,Thurner:multi, Menichetti, Halu}. Examples can be found everywhere. One is the multiplex consisting of the different social networks that a person may belong to. Other examples include 
the Internet's IPv4 and IPv6 topologies, or the structural and functional networks in the brain. Understanding the relations among the networks comprising a multiplex is crucial for understanding the behavior of a wide range of real world systems~\cite{our:model, ecology20, worldmodel, DeDomenico2014, Simas2015}. However, despite the burst of recent research in studying the properties of multiplex networks, e.g.,~\cite{arenas:multiplex,multilayer:kivel,Boccaletti20141}, a universal framework describing the relations among the single networks comprising a multiplex and what implications these relations may have when it comes to applications remains elusive.

Here, we show that real multiplexes are not random combinations of single network layers. Instead, we find that their constituent networks exhibit significant \emph{hidden geometric correlations}. These correlations are called ``hidden" as they are not directly observable by looking at each individual network topology. Specifically, each single network can be mapped (or embedded) into a separate hyperbolic space~\cite{Krioukov2010,Boguna2010,frag:hypermap,frag:hypermap_cn}, where node coordinates abstract the \emph{popularity} and \emph{similarity} of nodes~\cite{Serrano2008,boguna:popularity}. We find that node coordinates are significantly correlated across layers of real multiplexes, meaning that distances between nodes in the underlying hyperbolic spaces of the constituent networks are also significantly correlated.

The discovered geometric correlations yield a very powerful framework for answering important questions related to real multiplexes. Specifically, we first show that these correlations suggest the existence of multidimensional communities, which are sets of nodes that are similar (close in the underlying space) in multiple layers. Further, we show that strong geometric correlations imply accurate trans-layer link prediction, where connections in one layer can be predicted by knowing the hyperbolic distances among nodes in another layer. This is important for applications where we only know the connections among nodes in one context, e.g., structural connections between brain regions, and we want to predict connections between the same nodes in some other context, e.g., likelihood of functional connections between the same brain regions. 

Finally, to study the effects of geometric correlations on dynamical processes, we consider targeted navigation. Targeted navigation is a key function of many real networks, where either goods, people, or information is transferred from a source to a destination using the connections of the network. It has been shown that single complex networks, like the IPv4 Internet or the network of airport connections, can be navigated efficiently by performing \emph{greedy routing} in their underlying geometric spaces~\cite{Boguna2008, Boguna2010, Papadopoulos2010, Krioukov2010}. In greedy routing, nodes forward incoming messages to their neighbors that are closest to the destination in the geometric space so that they only need local knowledge about the coordinates of their neighbors. In multilayer systems, messages can switch between layers and so a node forwards a message to its neighbor that is closest to the destination in any of the layers comprising the system. We call this process \emph{mutual greedy routing} (MGR).

Mutual greedy routing follows the same line of reasoning as greedy routing in Milgram's experiment~\cite{milgram1969}, which was indeed performed using multiple domains.
For example, in the case of a single network, to reach a lawyer in Boston one might want to forward a message to a judge in Los Angeles (greedy routing). However, in the case of two network layers, it might be known that the lawyer in Boston is also a passionate vintage model train collector. An individual who knows a judge in Los Angeles and the owner of a vintage model train shop in New York would probably choose to forward the message to the latter (MGR). Similarly, air travel networks can be supported by train networks to enhance the possibilities to navigate the physical world, individuals can use different online social networks to increase their outreach, and so on. In this work, we consider the real Internet, which is used to navigate the digital world, and show that mutual greedy routing in the multiplex consisting of the IPv4 and IPv6 Internet topologies~\cite{as_topo} outperforms greedy routing in the single IPv4 and IPv6 networks. We also use synthetic model networks to show that geometric correlations improve the navigation of multilayer systems, which outperforms navigation in the single layers if these correlations are sufficiently strong. As we will show, if optimal correlations are present, the fraction of failed deliveries is mitigated superlinerarly with the number of layers, suggesting that more layers with the right correlations quickly make multiplex systems almost perfectly navigable. 

%%%%%%%%%%%%%%%%%%%%%%%%%%%%%%%%%%%%%%%%%%%%%%%%%%%%%%%%%%%%%%%%%%%%%%%%%%%%%%%%%%%
\section*{Geometric organization of real multiplexes}
It has been shown that many real (single layer) complex networks have an effective or hidden geometry underneath their observed topologies, which is hyperbolic rather than Euclidean~\cite{Krioukov2009, Krioukov2010, Boguna2010, boguna:popularity,Serrano2011}. In this work, we extend the hidden geometry paradigm to real multiplexes and prove that the coordinates of nodes in the different underlying spaces of layers are correlated.

\subsection*{Geometry of single layer networks}
Nodes of real single-layered networks can be mapped to points in the hyperbolic plane, such that each node $i$ has the polar coordinates, or hidden variables, $r_i, \theta_i$. The radial coordinate $r_i$ abstracts the node popularity. The smaller the radial coordinate of a node, the more popular the node is, and the more likely it attracts connections. The angular distance between two nodes, $\Delta \theta_{ij}=\pi-|\pi-|\theta_i-\theta_j||$, abstracts their similarity. The smaller this distance, the more similar two nodes are, and the more likely they are connected. The hyperbolic distance between two nodes, very well approximated by $x_{ij} \approx r_i+r_j+2\ln{\sin{(\Delta\theta_{ij}/2)}}$~\cite{Krioukov2010}, is then a single-metric representation of a combination of the two attractiveness attributes, radial popularity and angular similarity. The smaller the hyperbolic distance between two nodes, the higher is the probability that the nodes are connected, meaning that connections take place by optimizing trade-offs between popularity and similarity~\cite{boguna:popularity}. 

Techniques based on Maximum Likelihood Estimation for inferring the popularity and similarity node coordinates in a real network have been derived in~\cite{Boguna2010} and recently optimized in~\cite{frag:hypermap, frag:hypermap_cn}. It has been shown that through the constructed hyperbolic maps one can identify soft communities of nodes, which are groups of nodes located close to each other in the angular similarity space~\cite{Boguna2010, Serrano2011, boguna:popularity, soft:comm}; predict missing links with high precision~\cite{Serrano2011,frag:hypermap, frag:hypermap_cn}; and facilitate efficient greedy routing in the Internet, which can reach destinations with more than $90$\% success rate, following almost shortest network paths~\cite{Boguna2010, frag:hypermap, frag:hypermap_cn}.

%%%%%%%%%%%%%%%%%%%%%%%%%%%%%%%%%%%%%%%%%%%%%%%%%%%%%%%%%%%%%%%%%%%%%%%%%%%%%%%%%%%
\subsection*{Geometry of real multiplexes}
\begin{figure}[b]
\centering
\includegraphics[width=1\linewidth]{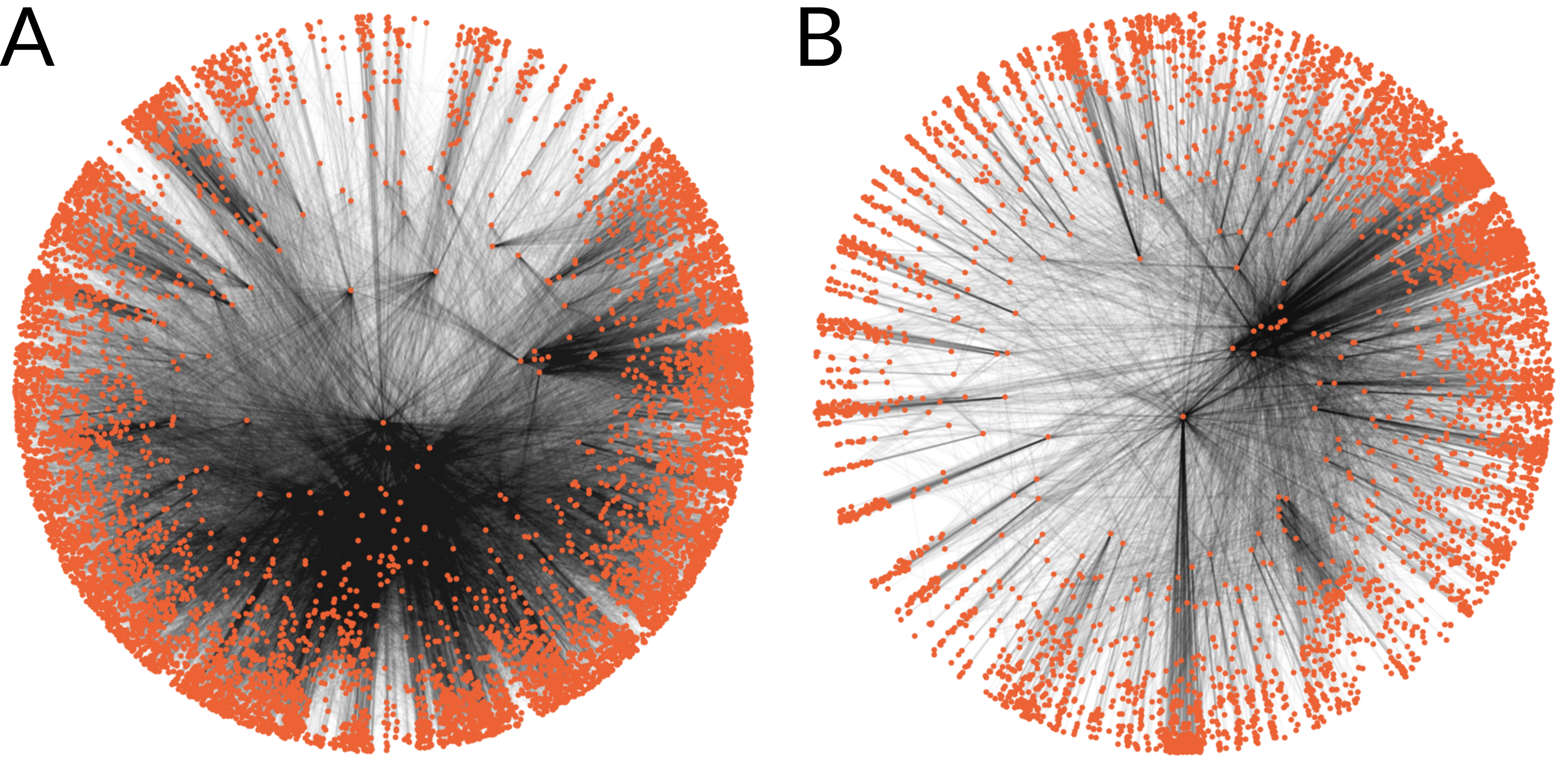}
\caption{\textbf{Hyperbolic mapping of the IPv4/IPv6 Internet.} \textbf{A:} IPv4 topology---for clarity only nodes with degrees greater than $3$ are shown. \textbf{B:} IPv6 topology. 
\label{fig_internet_embedding}}
\end{figure}
\begin{table*}[t]
\begin{ruledtabular}
\begin{tabular}{lllll}
\textbf{Name} &  \textbf{Type} & \textbf{Nodes} & \textbf{Layer~1, Layer~2, ...} &\textbf{Source} \\ \hline
Internet & Technological & Autonomous Systems & IPv4 AS topology, IPv6 AS topology &\cite{as_topo} \\ 
Air/Train & Technological & Airports/Train stations & Indian airport network, Indian train network &\cite{Halu} \\
Drosophila & Biological & Proteins & Suppressive genetic interaction, additive genetic interaction &\cite{biogrid,arenas:reduce} \\
C. Elegans & Biological & Neurons & Electric, chemical monadic, chemical polyadic synaptic junctions &\cite{pnas:celegans,muxviz} \\
Brain & Biological & Brain regions & Structural network, functional network &\cite{Simas2015} \\
SacchPomb & Biological & Proteins & 5 different interactions (direct interaction, colocalization, ...) & \cite{biogrid,arenas:reduce} \\
Rattus & Biological & Proteins & Physical association, direct interaction & \cite{biogrid,arenas:reduce}\\
arXiv & Collaboration & Authors &  8 different categories (physics.bio-ph, cond-mat.dis-nn, ...) & \cite{prx:modular} \\
Physicians & Social & Physicians &  Advice, discussion, friendship relations & \cite{physicians:data} 
\end{tabular}
\end{ruledtabular}
\caption{
Overview of the considered real-world multiplex network data. Details for each dataset can be found in Supplementary Information section~I.
\label{tab_datasets}}
\end{table*}

We now consider nine different real-world multiplex networks from diverse domains. An overview of the considered datasets is given in Table~\ref{tab_datasets}---see Supplementary Information section~I for a detailed description. For each multiplex, we map each network layer independently to an underlying hyperbolic space using the \emph{HyperMap} method~\cite{frag:hypermap, frag:hypermap_cn} (see Supplementary Information section~II), thus inferring the popularity and similarity coordinates $r,\theta$ of all of its nodes. A visualization of the mapped IPv4 and IPv6 Internet layers is shown in Fig.~\ref{fig_internet_embedding}.
For each of our real multiplexes, we find that both the radial and the angular coordinates of nodes that exist in different layers are correlated. 

The radial popularity coordinate of a node $i$ depends on its observed degree in the network $k_i$ via $r_i \sim \ln{N}-\ln{k_i}$, where $N$ is the total number of nodes~\cite{Boguna2010, frag:hypermap, frag:hypermap_cn}. Therefore, radial correlations are equivalent to correlations among node degrees, which have been recently found and studied~\cite{pre:multiplex:correlations, pre:degree:correlations:2, prl:degree:correlations, scirep:degree:corr, self:similar:multiplex}. Consistent with these findings, radial correlations are present in our real multiplexes and are encoded in the conditional probability $P(r_2|r_1)$ that a node has radial coordinate $r_2$ in layer 2 given its radial coordinate $r_1$ in layer 1. $P(r_2|r_1)$ for different real multiplexes is shown in Supplementary Fig.~1, where we observe significant radial correlations. 
\begin{figure*}[p]
\centering
\includegraphics[width=1\linewidth]{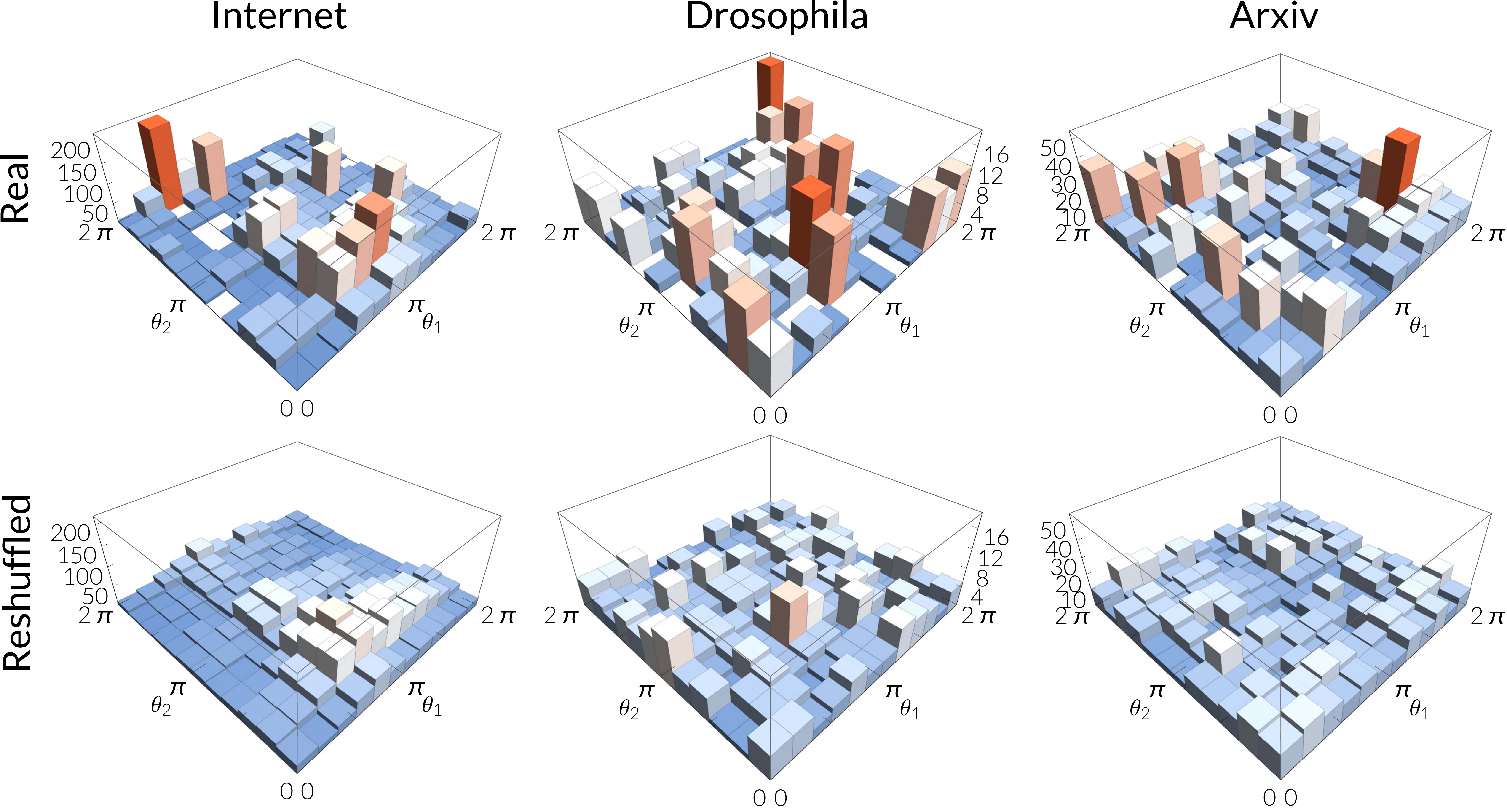}
\caption{\textbf{Distribution of nodes in the two-dimensional similarity space of the Internet, Drosophila, and arXiv multiplexes (top panel)}. The plots correspond to nodes that exist in both layers of each system. The angular similarity coordinate of a node in layer 1 is denoted by $\theta_1$ and in layer 2 by $\theta_2$. The histogram heights are equal to the number of nodes falling within each two-dimensional similarity bin, and the colors in each case denote the relative magnitude of the heights. \textbf{Bottom panel:} The same distributions as in the top panel but for the reshuffled counterparts of the real systems. 
\label{fig_histogram}}
\end{figure*}

The correlation among the angular similarity coordinates, on the other hand, is a fundamentally new result that has important practical implications. Fig.~\ref{fig_histogram} shows the distribution of nodes that have angular coordinates $(\theta_1,\theta_2)$ in layers $1$ and $2$ of the real Internet, Drosophila, and arXiv multiplexes. The figure also shows the corresponding distributions in the reshuffled counterparts of the real systems, where we have destroyed the trans-layer coordinate correlations by randomly reshuffling node ids (see  Supplementary Information section~III). We observe an overabundance of two-dimensional similarity clusters in the real multiplexes. These clusters consist of nodes that are similar, i.e., are located at small angular distances, in both layers of the multiplex. These similarity clusters do not exist in the reshuffled counterparts of the real systems, and are evidence of significant angular correlations. Similar results hold for the rest of the real multiplexes that we consider (see Supplementary Information section~IV). In Supplementary Information section~X, we quantify both the radial and the angular correlations present in all the considered real multiplexes.

The generalization of community definition and detection techniques from single layer networks to multiplex systems has recently gained attention~\cite{Mucha876, multiplex:community:cs, multiplex:communities, prx:modular}. The two-dimensional similarity histograms in Fig.~\ref{fig_histogram} allow one to naturally observe two-dimensional soft communities, defined as sets of nodes that are similar---close in the angular similarity space---in both layers of the multiplex simultaneously. They also provide a measure of distance between the different communities. We note that these communities are called ``soft" as they are defined by the geometric proximity of nodes~\cite{Boguna2010, Serrano2011, boguna:popularity, soft:comm} rather than by the network topology as is traditionally the case~\cite{Fortunato201075}. We also note that we do not develop an automated multidimensional community detection algorithm here. However, our findings could serve as a starting point for the development of such an algorithm (see discussion in Supplementary Information section~IV). In Supplementary Information section~V, we show how certain geographic regions are associated to two-dimensional communities in the IPv4/IPv6 Internet.

\begin{figure*}[p]
\centering
\includegraphics[width=01\linewidth]{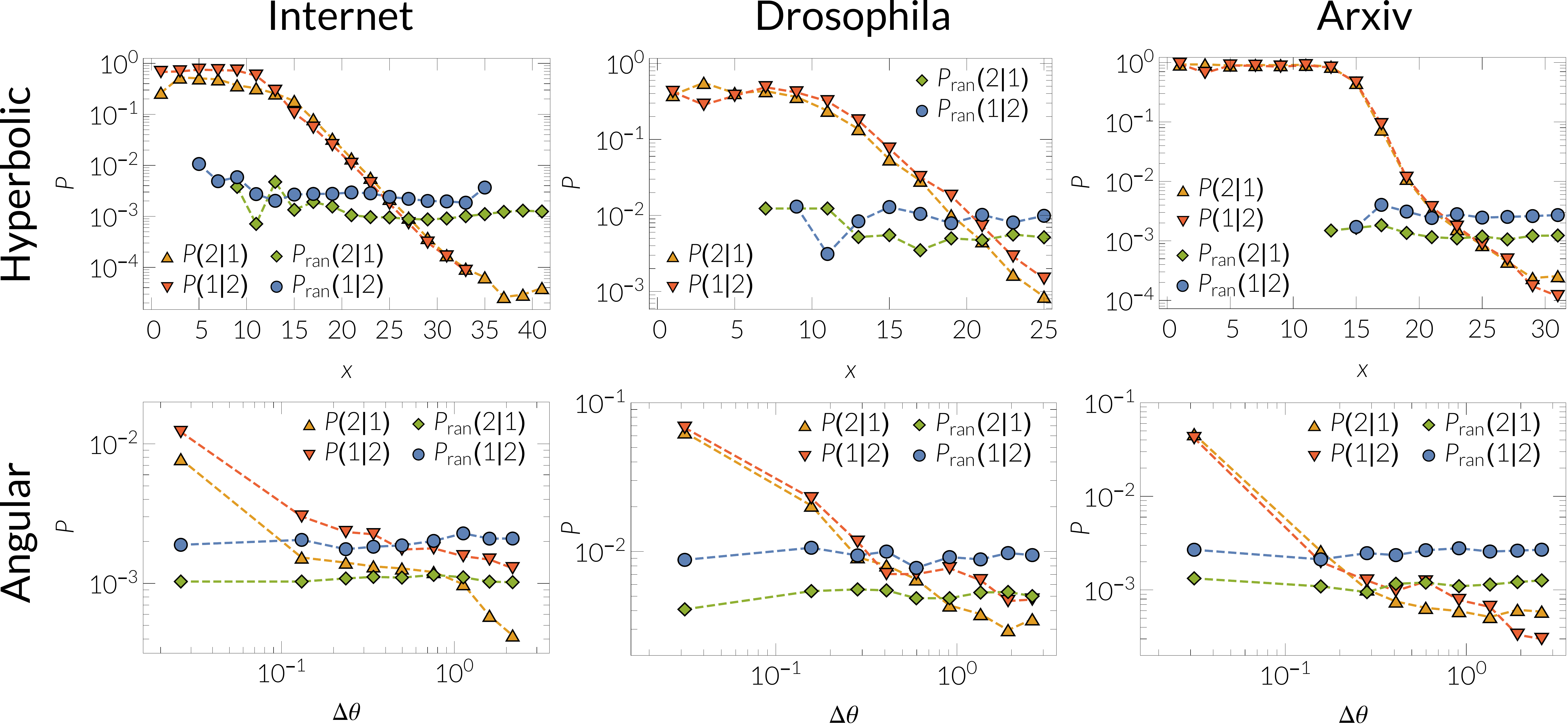}
\caption{\textbf{Trans-layer connection probability in the Internet, Drosophila, and arXiv multiplexes.} \textbf{Top panel:} Trans-layer connection probability as a function of hyperbolic distance. $P(j|i)$ denotes the probability that a pair of nodes is connected in layer $j$ given its hyperbolic distance $x$ in layer $i$. $P_{\text{ran}}(j|i)$ denotes the same probability for the reshuffled counterpart of each real system. \textbf{Bottom panel:} Corresponding trans-layer connection probabilities when considering only the angular (similarity) distance between nodes, $\Delta\theta$.
\label{fig_link_prediction}}
\end{figure*}

%%%%%%%%%%%%%%%%%%%%%%%%%%%%%%%%%%%%%%%%%%%%%%%%%%%%%%%%%%%%%%%%%%%%%%%%%%%%%%%%%%%
The problem of link prediction has been extensively studied in the context of predicting missing and future links in single layer networks~\cite{link:prediction:1,Clauset:2008:Hierarchicalstructure} and its generalization to real-world multilayer systems is recently gaining attention~\cite{multiplex:link:prediction:1}. In our case,
the radial and angular correlations across different layers suggest that by knowing the hyperbolic distance between a pair of nodes in one layer, we can predict the likelihood that the same pair is connected in another layer. Fig.~\ref{fig_link_prediction} (top) validates that this is indeed the case. The figure shows the empirical trans-layer connection probability, $P(1|2)$ ($P(2|1)$), that two nodes are connected in one of the layers of the multiplex given their hyperbolic distance in the other layer. In Fig.~\ref{fig_link_prediction} (top), we observe that this probability decreases with the hyperbolic distance between nodes in the considered real multiplexes (see Supplementary Information section~IV for the results for the rest of the multiplexes). By contrast, in their reshuffled counterparts, which do not exhibit geometric correlations, this probability is almost a straight line. Fig.~\ref{fig_link_prediction} (bottom) shows that the trans-layer connection probability decreases with the angular distance between nodes, which provides an alternative empirical validation of the existence of significant similarity correlations across the layers. In Supplementary Information section~VI, we quantify the quality of trans-layer link prediction in all the real multiplexes we consider, and show that our approach outperforms a binary link predictor that is based on edge overlaps.

%%%%%%%%%%%%%%%%%%%%%%%%%%%%%%%%%%%%%%%%%%%%%%%%%%%%%%%%%%%%%%%%%%%%%%%%%%%%%%%%%%%
\section*{Modeling geometric correlations and implications to mutual greedy routing}

The IPv4 Internet has been found to be navigable~\cite{Boguna2010, frag:hypermap, frag:hypermap_cn}. Specifically, it has been shown that greedy routing (GR) could reach destinations with more than 90\% success rate in the constructed hyperbolic maps of the IPv4 topology in 2009. We find a similar efficiency of GR in both the IPv4 and IPv6 topologies considered here, which correspond to January 2015. Specifically,  we perform GR in the hyperbolic map of each topology among $10^5$ randomly selected source-destination pairs that exist in both topologies. We find that GR reaches destinations with $90\%$ and $92\%$ success rates in IPv4 and IPv6, respectively. Furthermore, we also perform angular GR, which is the same as GR but using only angular distances. We find that the success rate in this case is almost $60\%$ in both the IPv4 and IPv6 topologies. However, hyperbolic mutual greedy routing (MGR) between the same source-destination pairs, which travels to any neighbor in any layer closer to destination, increases the success rate to $95\%$, while the angular MGR that uses only the angular distances, increases the success rate along the angular direction to $66\%$. More details about the MGR process are found in Supplementary Information section~XI. 

The observations above raise the following fundamental questions. (i) How do the radial and angular correlations affect the performance of MGR? (ii) Under which conditions does MGR perform better than single-layer GR? (iii) How does the performance of MGR depend on the number of layers in a multilayer system? And (iv), how close to the optimal---in terms of MGR's performance---are the geometric correlations in the IPv4/IPv6 Internet? Answering these questions requires a framework to construct realistic synthetic topologies (layers) where correlations---both radial and angular---can be tuned without altering the topological characteristics of each individual layer. 

\begin{figure*}[t]
\includegraphics[width=1\linewidth]{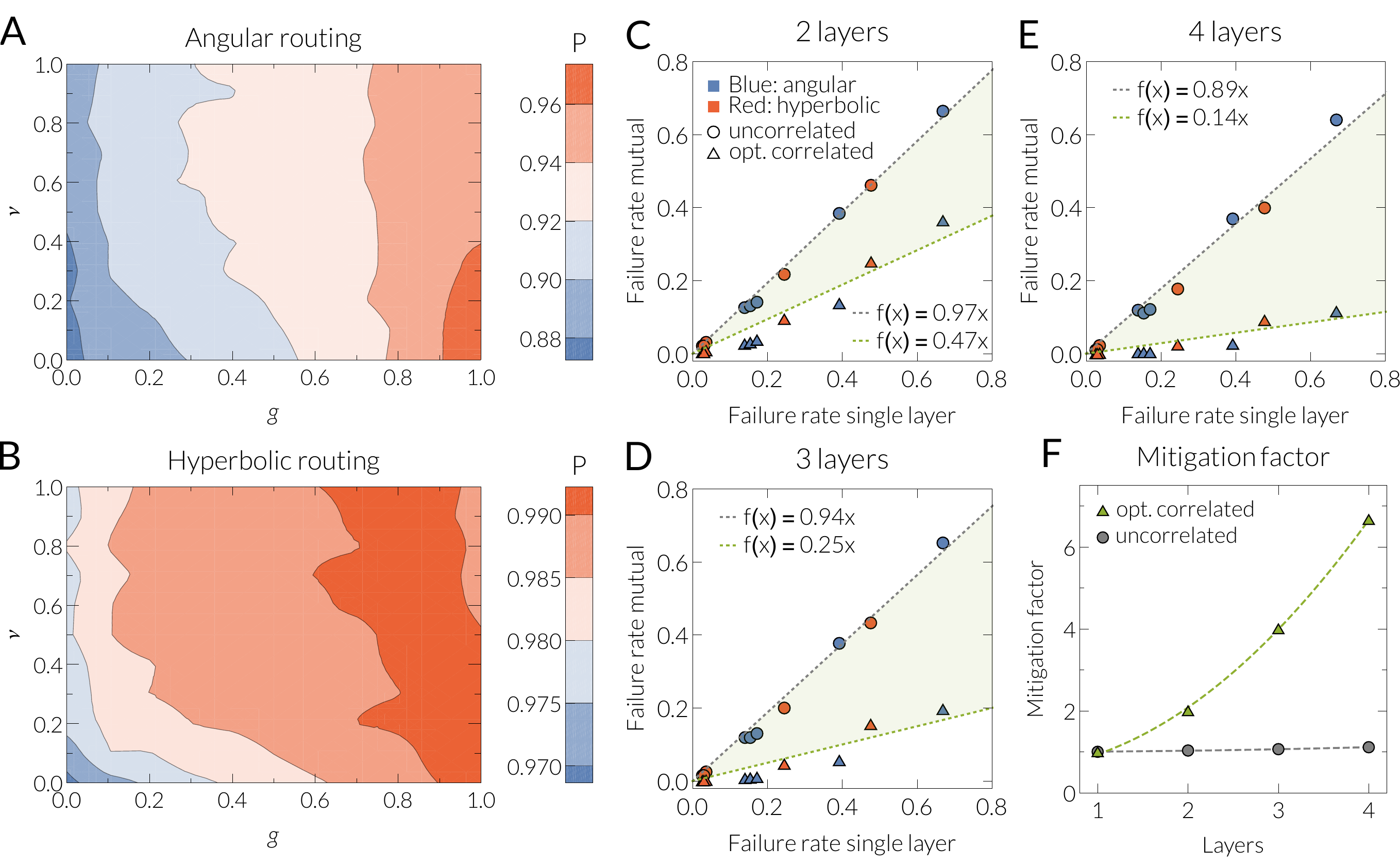}  
\caption{\textbf{Performance of mutual navigation in synthetic multiplexes with geometric correlations.} \textbf{A,~B:}~Success rate of angular MGR and of MGR for a two-layer multiplex system as a function of the radial ($\nu$) and angular ($g$) correlation strengths. Each layer has $N=30000$ nodes, power law degree distribution $P(k) \sim k^{-\gamma}$ with $\gamma=2.5$, average degree $\bar{k}=10$, and temperature $T=0.4$ (temperature controls clustering in the network, see Methods section~B). \textbf{C-E:}~Failure rate ($1-$success rate) of MGR (red triangles or circles) and angular MGR (blue triangles or circles). Each layer has the same parameters as in plots \textbf{A,~B} and the same temperature $T$ that takes different values, $T =(0.1, 0.2, 0.4, 0.8, 0.9)$, corresponding for each navigation type respectively to the triangle/circle points, from the leftmost triangle/circle to the rightmost triangle/circle. Circles correspond to the case where there are no coordinate correlations among the layers, while triangles correspond to the case where there are optimal correlations, i.e., radial and angular correlation strengths that maximize the corresponding performance of MGR or angular MGR. \textbf{F:}~Mitigation factor as a function of the number of layers for optimal coordinate correlations and for uncorrelated coordinates.
\label{fig_layers_panel}}
\end{figure*}

To this end, we develop the geometric multiplex model (GMM) (see Methods section B). The model generates multiplexes with layer topologies embedded in the hyperbolic plane and where correlations can be tuned independently by varying the model parameters $\nu \in [0,1]$ (radial correlations) and $g \in [0,1]$ (angular correlations). We consider two-, three- and four-layer multiplexes constructed using our framework with different values of the correlation strength parameters $\nu$ and $g$.

From Figs.~\ref{fig_layers_panel}A,B and Supplementary Figs.~16--18, we observe that in general both MGR and angular MGR perform better as we increase the correlation strengths $\nu$ and $g$. When both radial and angular correlations are weak, we do not observe any significant benefits from mutual navigation. Indeed, in Figs.~\ref{fig_layers_panel}C-E we observe that in the uncorrelated case ($\nu \to 0$, $g\to 0$) MGR performs almost identical to the single-layer GR, irrespectively of the number of layers. This is because when a message reaches a node in one layer after the first iteration of the MGR process, the probability that this node will have a neighbor in another layer closer to the destination is small. That is, even though a node may have more options (neighbors in other layers) for forwarding a message, these options are basically useless and messages navigate mainly single layers. 

Increasing the strength of correlations makes MGR more efficient, as the probability to have a neighbor closer to the destination in another layer increases. However, increasing the strength of correlations also increases the edge overlap between the layers (see Supplementary Information section~VIII), which reduces the options that a node has for forwarding a message. We observe that very strong radial and angular correlations may not be optimal at low temperatures 
(cf. Supplementary Figs.~16--18 for $T=0.1$). This is because when the layers have the same nodes and the same average degree and power law exponent, $\bar{k}, \gamma$, then as $\nu \to 1$, $g \to 1$, the coordinates of the nodes in the layers become identical (see Supplementary Information section~VII). Further, if the temperature of the layers is low, the connection probability $p(x_{ij})$ in each layer becomes the step function, where two nodes $i,j$ are deterministically connected if their hyperbolic distance is $x_{ij} \leq R~\sim \ln{N}$ (see Methods section~B). That is, as $T \to 0$, $\nu \to 1$, $g \to 1$, all layers become identical, and MGR degenerates to single-layer GR. We observe (Figs.~\ref{fig_layers_panel}A,B and Supplementary Figs.~16--18) that the best MGR performance is always achieved at high angular correlations, and either high radial correlations if the temperature of the individual layers is high, or low radial correlations if the temperature of the layers is low. The best angular MGR performance is always achieved at high angular and low radial correlations. In other words, MGR performs more efficient when the layers comprising the multiplex are similar but not ``too'' similar. High temperatures induce more randomness so that even with maximal geometric correlations the layers are not too similar. However, at low temperatures, very strong geometric correlations make the layers too similar and the best performance occurs for intermediate correlations.

From Figs.~\ref{fig_layers_panel}C-E we observe that, for a fixed number of layers and for optimal correlations, the failure rate ($1-$success rate) is reduced by a constant factor, which is independent of the navigation type (MGR or angular MGR) and the layer temperature. This factor, which we call \emph{failure mitigation factor}, is the inverse of the slope of the best-fit lines in Figs.~\ref{fig_layers_panel}C-E. Fig.~\ref{fig_layers_panel}F shows the failure mitigation factor for our two-, three- and four-layer multiplexes for both uncorrelated and optimally correlated coordinates.  Remarkably, if optimal correlations are present, the failure mitigation factor grows superlinerarly with the number of layers, suggesting that more layers with the right correlations can quickly make multiplex systems almost perfectly navigable. On the contrary, more layers without correlations do not have a significant effect on mutual navigation, which performs virtually identical to single-layer navigation.

%%%%%%%%%%%%%%%%%%%%%%%%%%%%%%%%%%%%%%%%%%%%%%%%%%%%%%%%%%%%%%%%%%%%%%%%%%%%%%%%%%%

\begin{figure}[t]
\includegraphics[width=1\linewidth]{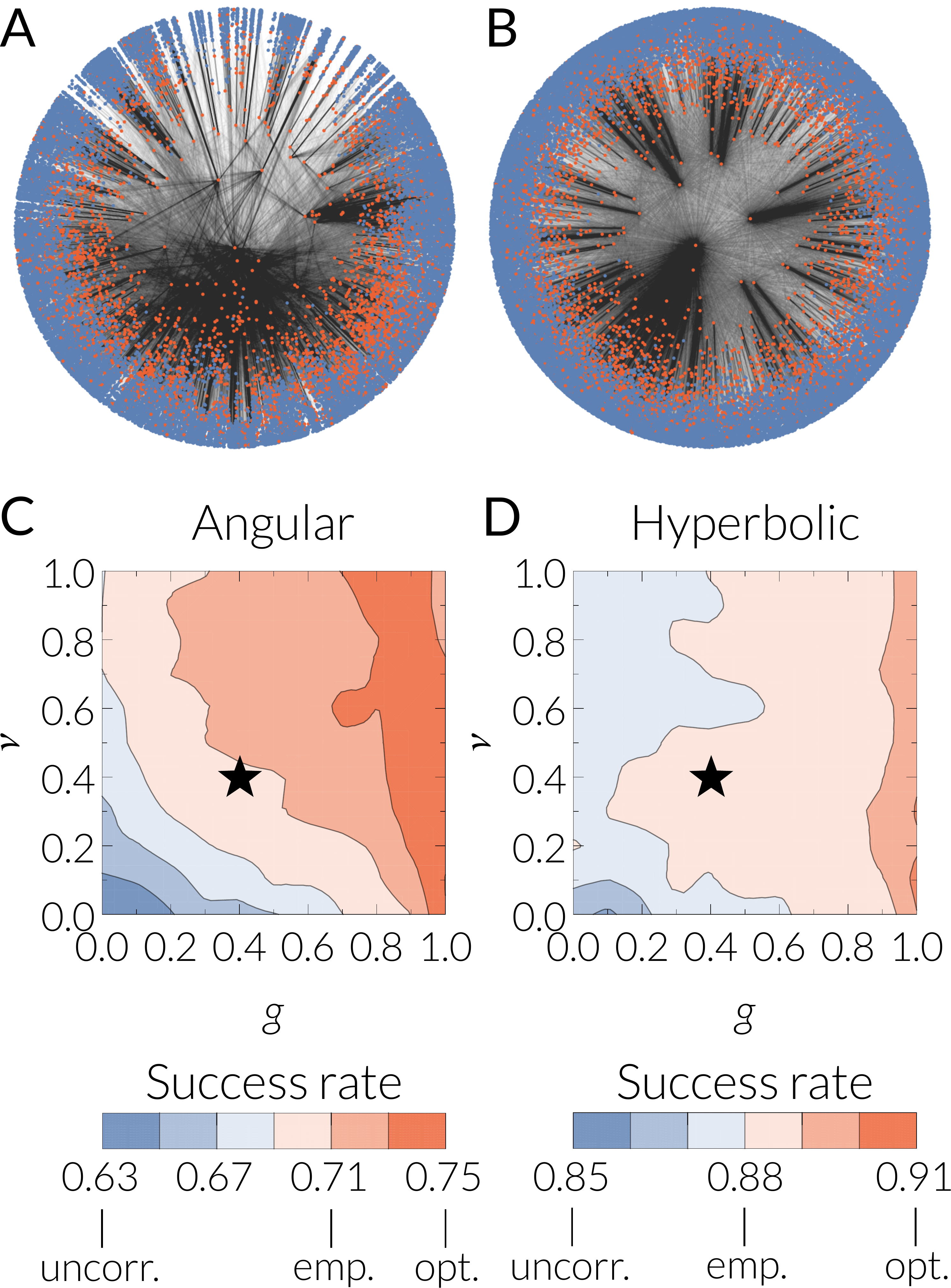}
\caption{\textbf{Performance of mutual navigation as a function of radial and angular correlation strengths ($\nu, g$) in a two-layer synthetic multiplex that best mimics the real IPv4/IPv6 Internet.} \textbf{A:}~Hyperbolic mapping of the real IPv4 topology where nodes marked by red also exist in the IPv6 topology. \textbf{B:}~Hyperbolic mapping of layer 1 of our Internet-like synthetic multiplex, where nodes marked by red also exist in layer 2. \textbf{C}:~Performance of angular MGR. \textbf{D}:~Performance of MGR. In \textbf{C, D} the black star indicates the achieved performance with the estimated correlation strengths in the real IPv4/IPv6 Internet, $\nu_E \approx 0.4, g_E \approx 0.4$.
 \label{fig_psi_fitted_graphs}} 
\end{figure}
Finally, we investigate how close to the optimal---in terms of mutual navigation performance---are the radial and angular correlations in the IPv4/IPv6 Internet. To this end, we use an extension of the GMM framework that can account for different layer sizes (see Methods section~C) to construct an Internet-like synthetic multiplex. For those nodes that exist in both layers of our multiplex (see Figs.~\ref{fig_psi_fitted_graphs}A,B), we tune the correlations among their coordinates as before, by varying the parameters $\nu$ and $g$, and perform mutual navigation. Figs.~\ref{fig_psi_fitted_graphs}C,D show the performance of angular MGR and of MGR, respectively. We observe again that increasing the correlation strengths improves performance. In angular MGR, the success rate is $63\%$ with uncorrelated coordinates, while with optimal correlations it becomes $75\%$. In MGR, the success rate with uncorrelated coordinates is $85\%$ and with optimal correlations is $91\%$. The star in Figs.~\ref{fig_psi_fitted_graphs}C,D indicates the achieved performance with the empirical correlation strengths in the IPv4/IPv6 Internet, $\nu_E \approx 0.4, g_E \approx 0.4$, which are estimated using the inferred radial and angular coordinates of nodes (see Supplementary Information section~IX). At $\nu=\nu_E, g=g_E$, the success rate of angular MGR is $71\%$, which is closer to the rate obtained with optimal correlations than to the uncorrelated case. For MGR, the success rate is $88\%$, which lies in the middle of the uncorrelated and optimally correlated cases.

%%%%%%%%%%%%%%%%%%%%%%%%%%%%%%%%%%%%%%%%%%%%%%%%%%%%%%%%%%%%%%%%%%%%%%%%%%%%%%%%%%%
\section*{Outlook}

Hidden metric spaces underlying real complex networks provide a fundamental explanation of their observed topologies and, at the same time, offer practical solutions to many of the challenges that the new science of networks is facing nowadays. On the other hand, the multiplex nature of numerous real-world systems has proven to induce emerging behaviors that cannot be observed in single layer networks. The discovery of hidden metric spaces underlying real multiplexes is, thus, a natural step forward towards a fundamental understanding of such systems. The task is, however, complex due to the inherent complexity of each single layer and that, in principle, the metric spaces governing the topologies of the different single networks could be unrelated. Our results here clearly indicate that this is not the case.
  
Our findings open the door for multiplex embedding, in which all the layers of a real multiplex are simultaneously and not independently embedded into hyperbolic spaces. This would require a reverse engineering of the geometric multiplex model proposed here, which generates multiplexes where radial and angular correlations can be tuned independently by varying its parameters. Multiplex embeddings can have important applications in diverse domains, ranging from improving information transport and navigation or search in multilayer communication systems and decentralized data architectures~\cite{Contreras11122015,helbing:digital_democracy}, to understanding functional and structural brain networks and deciphering their precise relationship(s)~\cite{Simas2015}, to predicting links among nodes (e.g., terrorists) in a specific network by knowing their connectivity in some other network.

%%%%%%%%%%%%%%%%%%%%%%%%%%%%%%%%%%%%%%%%%%%%%%%%%%%%%%%%%%%%%%%%%%%%%%%%%%%%%%%%%%%
\section*{Methods}

\subsection{Computation of trans-layer connection probability}

To compute the trans-layer connection probability, we consider all nodes that exist in both layers. In each of the layers, we bin the range of hyperbolic distances between these nodes from zero to the maximum distance into small bins. For each bin we then find all the node pairs located at the hyperbolic distances falling within the bin. The percentage of pairs in this set of pairs that are connected by a link in the other layer, is the value of the empirical trans-layer connection probability at the bin. 

%%%%%%%%%%%%%%%%%%%%%%%%%%%%%%%%%%%%%%%%%%%%%%%%%%%%%%%%%%%%%%%%%%%%%%%%%%%%%%%%%%%
\subsection{Geometric multiplex model (GMM)}

Our framework builds on the (single-layer) network construction procedure prescribed by the newtonian $\mathbb{S}^{1}$~\cite{Serrano2008} and hyperbolic $\mathbb{H}^{2}$~\cite{Krioukov2010} models. The two models are isomorphic and here we present the results for the $\mathbb{H}^{2}$ version even if for calculations it is more convenient to make use of the $\mathbb{S}^{1}$. We recall that to construct a network of size $N$, the $\mathbb{H}^{2}$ model firsts assigns to each node $i=1,\ldots, N$ its popularity and similarity coordinates $r_i, \theta_i$. Subsequently, it connects each pair of nodes $i, j$ with probability $p(x_{ij})=1/(1+e^{\frac{1}{2T}(x_{ij}-R)})$, where $x_{ij}$ is the hyperbolic distance between the nodes and $R \sim \ln{N}$ (see Supplementary Information section~VII~A). The connection probability $p(x_{ij})$ is nothing but the Fermi-Dirac distribution. Parameter $T$ is the \emph{temperature} and controls clustering in the network~\cite{Dorogovtsev10-book}, which is the probability that two neighbors of a node are connected. The average clustering $\bar{c}$ is maximized at $T=0$, nearly linearly decreases to zero with $T \in [0,1)$, and is asymptotically zero if $T>1$. As $T \to 0$ the connection probability becomes the step function $p(x_{ij}) \to 1$ if $x_{ij} \leq R$, and $p(x_{ij}) \to 0$ if $x_{ij} > R$. It has been shown that the $\mathbb{S}^{1}$ and $\mathbb{H}^{2}$ models can construct synthetic networks that resemble real networks across a wide range of structural characteristics, including power law degree distributions and strong clustering~\cite{Serrano2008,Krioukov2010}.  Our framework constructs single-layer topologies using these models, and allows for radial and angular coordinate correlations across the different layers. The strength of these correlations can be tuned via model parameters $\nu \in [0,1]$ and $g \in [0,1]$, without affecting the topological characteristics of the individual layers, which can have different properties and different sizes (see Supplementary Information section~VII). The radial correlations increase with parameter $\nu$---at $\nu=0$ there are no radial correlations, while at $\nu=1$ radial correlations are maximized. Similarly, the angular correlations increase with parameter $g$---at $g=0$ there are no angular correlations, while at $g=1$ angular correlations are maximized. 

%%%%%%%%%%%%%%%%%%%%%%%%%%%%%%%%%%%%%%%%%%%%%%%%%%%%%%%%%%%%%%%%%%%%%%%%%%%%%%%%%%%
\subsection{Extended GMM and the IPv4/IPv6 Internet}

To construct the Internet-like synthetic multiplex of Fig.~\ref{fig_psi_fitted_graphs} we use an extention of the GMM that can construct multiplexes with different layer sizes and where correlations among the coordinates of the common nodes between the layers can be tuned as before via parameters $\nu$ and $g$ (see Supplementary Information section~VII~D).

Layer 1 in the Internet-like synthetic multiplex has approximately the same number of nodes as in the IPv4 topology, $N_1=37563$ nodes, as well as the same power law degree distribution exponent $\gamma_1=2.1$, average node degree $\bar{k}_1\approx 5$, and average clustering $\bar{c}_1 \approx 0.63$ ($T_1=0.5$). Layer 2 has approximately the same number of nodes as in the IPv6 topology, $N_2=5163$ nodes, and the same power law exponent $\gamma_2=2.1$, average node degree $\bar{k}_2\approx 5.2$, and average clustering $\bar{c}_2 \approx 0.55$ ($T_2=0.5$).

The IPv4 topology is significantly larger than the IPv6 topology, and there are $4819$ common nodes (Autonomous Systems) in the two topologies. We find that nodes with a higher degree in IPv4 are more likely to also exist in IPv6. Specifically, we find that the empirical probability $\psi(k)$ that a node of degree $k$ in IPv4 also exists in IPv6 can be approximated by $\psi(k)=1/(1+15.4k^{-1.05})$ (see Supplementary Fig.~10).
We capture this effect in our synthetic multiplex by first constructing layer 1, and then sampling with the empirical probability $\psi(k)$ nodes from layer 1 that will also be present in layer 2 (see Supplementary Information section~VII~D). A visualization illustrating the common nodes in the real Internet and in our synthetic multiplex is given in Figs.~\ref{fig_psi_fitted_graphs}A,B. We note that the fact that nodes with higher degrees in the larger layer have higher probability to also exist in the smaller layer has also been observed in several other real multiplexes~\cite{pre:multiplex:correlations}. However, our model for constructing synthetic multiplexes with different layer sizes  is quite general, and allows for any sampling function $\psi(k)$ to be applied (see Supplementary Information section~VII~D).
%%%%%%%%%%%%%%%%%%%%%%%%%%%%%%%%%%%%%%%%%%%%%%%%%%%%%%%%%%%%%%%%%%%%%%%%%%%%%%%%%%%

%

%%%%%%%%%%%%%%%%%%%%%%%%%%%%%%%%%%%%%%%%%%%%%%%%%%%%%%%%%%%%%%%%%%%%%%%%%%%%%%%%%%%

\noindent {\bf Acknowledgements}\\
This work was supported by: the European Commission through the Marie Curie ITN ``iSocial'' grant no.\ PITN-GA-2012-316808; a James S. McDonnell Foundation Scholar Award in Complex Systems; the ICREA Academia prize, funded by the {\it Generalitat de Catalunya}; the MINECO project no.\ FIS2013-47282-C2-1-P;  and the {\it Generalitat de Catalunya} grant no.\ 2014SGR608. Furthermore, M.~B. and M. A. S. acknowledge support from the European Commission FET-Proactive Project  MULTIPLEX no.\ 317532.

\end{document}